\documentclass[twocolumn,superscriptaddress,notitlepage, nofootinbib]{revtex4-1}

\usepackage{graphicx}
\usepackage{amsmath}
\usepackage{hyperref}
\usepackage[utf8]{inputenc}
\usepackage[dvipsnames]{xcolor}
\usepackage{subfigure}
\usepackage{caption}

\usepackage{dcolumn}

\hypersetup{colorlinks,linkcolor={blue},citecolor={teal},urlcolor={violet}}

\newcommand{\GRAPPA}{GRAPPA Institute,
University of Amsterdam, 1098 XH Amsterdam, The Netherlands}
\newcommand{\KAVLI}{Kavli Institute for the Physics and Mathematics of the Universe (Kavli IPMU, WPI), University of Tokyo,
Kashiwa, Chiba 277-8583, Japan}

\begin{document}

\title{Warm Dark Matter Constraints Using Milky-Way Satellite Observations and Subhalo Evolution Modeling}

\author{Ariane~Dekker}
\email{a.h.dekker@uva.nl}
\affiliation{\GRAPPA}

\author{Shin’ichiro~Ando}
\email{s.ando@uva.nl}
\affiliation{\GRAPPA}
\affiliation{\KAVLI}

\author{Camila~A.~Correa}
\email{c.a.correa@uva.nl}
\affiliation{\GRAPPA}

\author{Kenny~C.~Y.~Ng}
\email{kcyng@cuhk.edu.hk}
\affiliation{Department of Physics, The Chinese University of Hong Kong, Shatin, Hong Kong China}

\date{\today}

\begin{abstract}
Warm dark matter (WDM) can potentially explain small-scale observations that currently challenge the cold dark matter (CDM) model, as warm particles suppress structure formation due to free-streaming effects. Observing small-scale matter distribution provides a valuable way to distinguish between CDM and WDM. 
In this work, we use observations from the Dark Energy Survey and PanSTARRS1, which observe 270 Milky-Way satellites after completeness corrections. We test WDM models by comparing the number of satellites in the Milky Way with predictions derived from the Semi-Analytical SubHalo Inference ModelIng (\texttt{SASHIMI}) code, which we develop based on the extended Press-Schechter formalism and subhalos' tidal evolution prescription. We robustly rule out WDM with masses lighter than 4.4~keV at 95\% confidence level for the Milky-Way halo mass of $10^{12} M_\odot$. 
The limits are a weak function of the (yet uncertain) Milky-Way halo mass, and vary as $m_{\rm WDM}\agt 3.6$--$5.1$~keV for $(0.6$--$2.0) \times 10^{12} M_\odot$. For the sterile neutrinos that form a subclass of WDM, we obtain the constraints of $m_{\nu_s}>12$~keV for the Milky-Way halo mass of $10^{12} M_{\odot}$, independent of the mixing angle. These results based on \texttt{SASHIMI} do not rely on any assumptions of galaxy formation physics or are not limited by numerical resolution. The models, therefore, offer a robust and fast way to constrain the WDM models. By applying a satellite forming condition, however, we can rule out the WDM mass lighter than 9.0~keV for the Milky-Way halo mass of $10^{12} M_\odot$. 
\end{abstract}

\maketitle

\section{Introduction}
In the standard cold dark matter (CDM) cosmological model, structure forms through hierarchical merging of dark matter halos. CDM explains the observed structure well above $\sim$1~Mpc, whereas it has issues explaining small-scale observations~\cite{Bullock_2017} such as the missing satellite~\cite{Klypin:1999uc,Moore:1999nt}, core-cusp~\cite{Flores_1994,cite-key,de_Vega_2014}, and the too-big-to-fail~\cite{Boylan-Kolchin:2011qkt,Parry_2011} problems. Galaxy formation physics could help solving these issues~\cite{Sawala_2016,DiCintio:2014xia}, but is not enough as it cannot explain the over-prediction of the most massive subhalos~\cite{Boylan-Kolchin:2011qkt,Boylan_Kolchin_2012}. 
Alternatively, dark matter could be made out of warm dark matter that suppresses structure on small scales, while behaving as CDM on large scales being consistent with the cosmic-microwave-background and large-scale-structure observations~\cite{Anderson_2012,2014}. Sterile neutrinos are one of the well motivated WDM candidates~\cite{PhysRevLett.82.2832,PhysRevD.64.023501,Dolgov:2000ew,Canetti:2012kh}. 

Observing small-scale structure thus provides a valuable way to distinguish between CDM and WDM models, and indeed, various observations of strong gravitational lensing~\cite{Vegetti_2018,Hsueh_2019,Gilman_2019} and Lyman-$\alpha$ forests~\cite{McQuinn_2016,Ir_i__2017,Enzi_2021} were adopted to set constraints on the WDM mass of $>5.58$~keV and $>3.5$~keV, respectively. As these constraints depend on different observations and assumptions, it is important to have complementary searches. 
Another excellent probe to test the WDM models are the satellite galaxies in the Milky Way, as the abundance of these systems are suppressed for WDM models. Satellite galaxies are formed through complex astrophysical processes within dark matter subhalos~\cite{Wechsler_2018,Somerville_2015}, which are smaller halos that accreted onto a larger host. 

Subhalo properties can be well estimated using cosmological N-body simulations~\cite{Lovell_2014,Tormen:1996fc,Gao_2004}. They are, however, limited by numerical resolution, motivating towards accurate analytical and semi-analytical models. Indeed, in order to test WDM models by studying Milky-Way satellite counts, previous studies have discussed semi-analytical models, based on the extended Press-Schechter (EPS) formalism~\cite{schneider2015structure,Cherry_2017}. The EPS formalism provides analytical expressions for the hierarchical assembly of dark matter halos, where halos are formed through gravitational collapse of a density fluctuation above a critical value~\cite{Press:1973iz,Bond:1990iw,Lacey:1993iv}. 
Not all subhalos host satellites and Refs.~\cite{Newton_2021,Nadler_2021,Kennedy_2014,Escudero_2018} adopt a galaxy formation model for the galaxy-halo connection, while Refs.~\cite{schneider2015structure,Cherry_2017} adopt a threshold on the halo mass above which star formation is initiated. These are, however, model-dependent and the results are affected by the choice of model parameters. 

In this work, we present a semi-analytical model based on the EPS, combined with semi-analytical relations that describe the halo and subhalo evolution. Subhalos lose mass through gravitational tidal stripping after they accrete onto their host, changing the internal structure as well as completely disrupting subhalos within certain radii~\cite{jiang2014statistics,vandenBosch:2004zs,Giocoli_2008}. This has not been taken into account in any previous semi-analytical work with WDM models, while it has been described for the case of CDM in Ref.~\cite{Hiroshima_2018}. 
In this work, we build on the semi-analytical models of Ref.~\cite{Hiroshima_2018} and extend it to the WDM cosmology by modifying the mass-loss rate, and adopting appropriate changes to the EPS formalism~\cite{Benson_2012} and to the concentration-mass-redshift relation for WDM~\cite{Ludlow_2016}. Our models enable us to directly probe subhalo properties for any WDM models as well as any halo and subhalo masses, resulting in competent and solid constraints, for which we make extensive comparison pointing out differences among various approaches. 

We calculate the number of satellite galaxies in the Milky Way for a range of WDM and sterile neutrino models and compare them with the observed number of satellite galaxies. For observational data, we use 270 estimated satellite galaxies observed by the Dark Energy Survey (DES) and PanSTARRS1 (PS1) after completeness correction~\cite{Drlica_Wagner_2020}, as well as a subset of 94 satellite galaxies that contain kinematics data, to obtain lower limits on the WDM and sterile neutrino mass. 
To derive our canonical, {\it conservative} constraints, we assume that all the subhalos host satellite galaxies. 
Implementing galaxy formation in subhalos above some certain thresholds (such as mass) will effectively reduce the number of satellites that the models predict and lead to stronger limits. Therefore, we also investigate different galaxy formation conditions. 

\begin{figure}[ht!]
    \centering
    \hskip7.mm
    \includegraphics[width=0.48\textwidth]{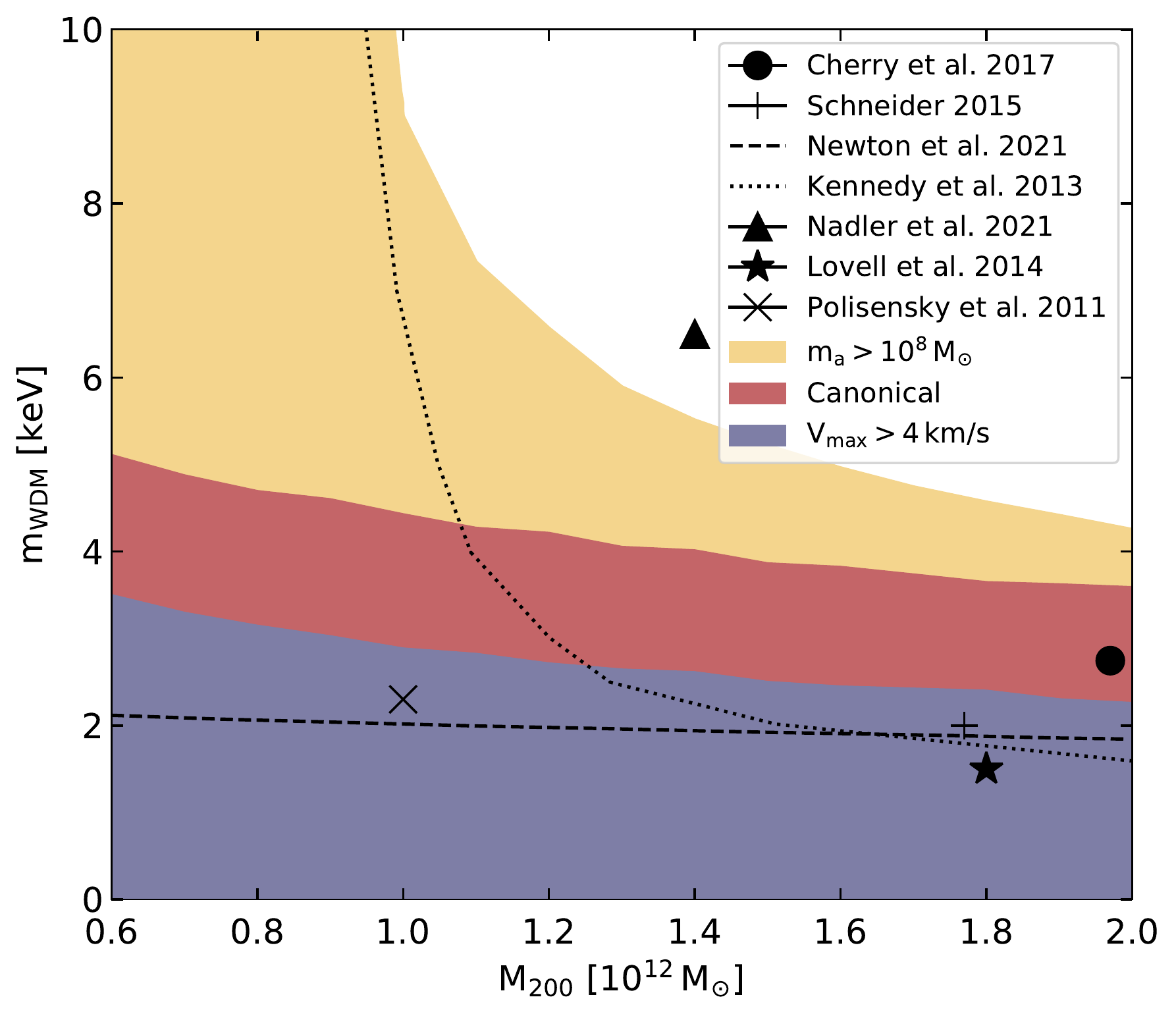}
\caption{Excluded regions at 95\% CL of the WDM mass as a function of the Milky-Way mass considering the canonical constraints (red) as well as by adopting the satellite forming condition with $m_a>10^8M_{\odot}$ (yellow). Moreover, the conservative constraints considering satellites with kinematics data of $V_{\rm max}>4$~km/s are also shown (purple). The black markers represent limits from the literature (Sec.~\ref{sec:disc}).}
    \label{fig:wdm}
\end{figure}

\begin{figure}[ht!]
    \centering
    \hskip7.mm
    \includegraphics[width=0.48\textwidth]{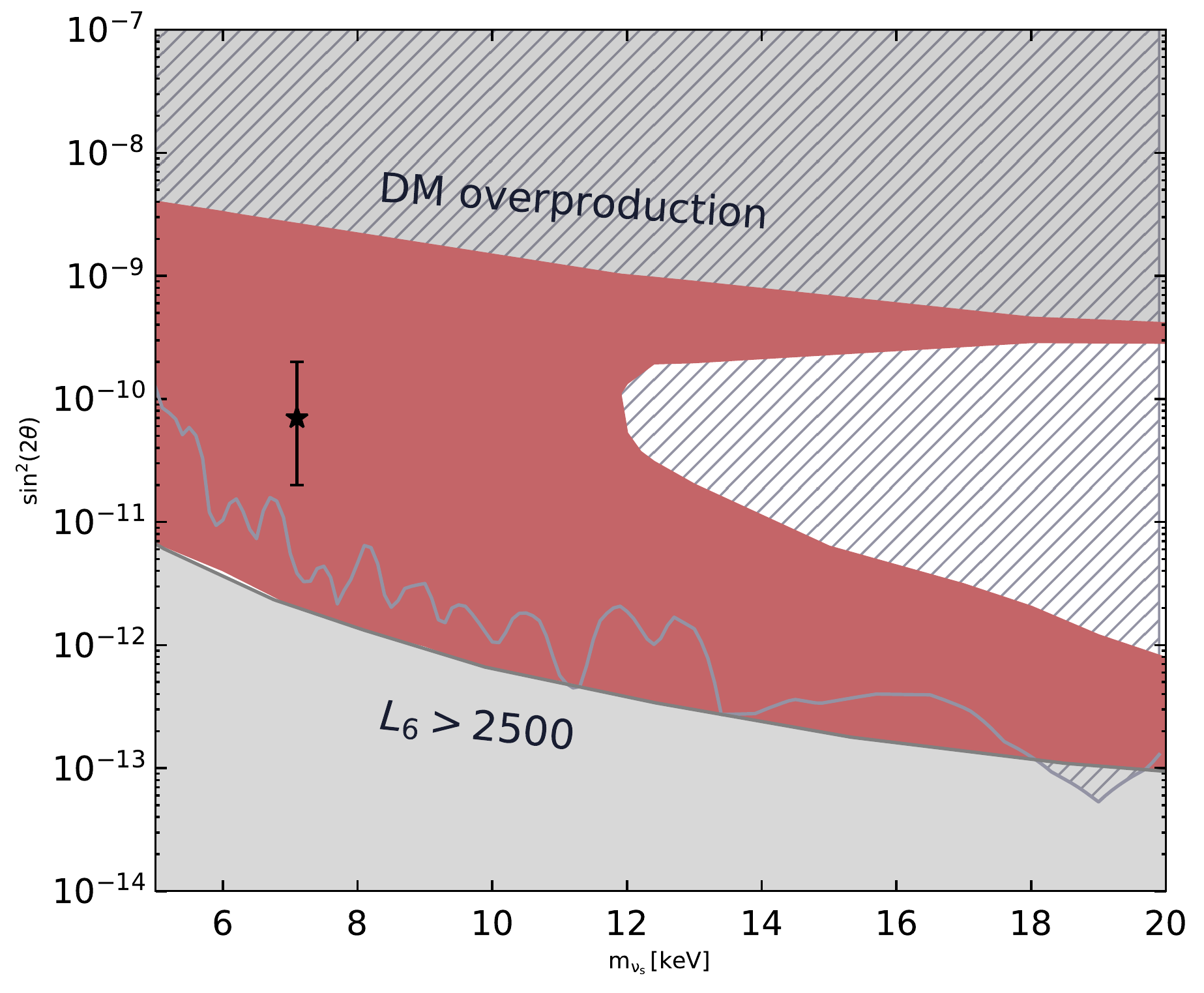}
\caption{Excluded regions at 95\% CL of the mixing angle $\sin^2(2\theta)$ as a function of sterile neutrino mass $m_{\nu_s}$ for the Milky-Way mass of $M_{200}=10^{12}M_{\odot}$. The grey hatched area represents upper limits from the current X-ray constraints~\cite{Horiuchi:2013noa, Ng:2015gfa, Perez:2016tcq, Ng:2019gch,Abazajian_2017,Caputo:2019djj,Roach_2020,foster2021deep} and the black star the best-fit of the unidentified $3.5$~keV line with mixing angle, $\sin^2(2\theta)\simeq (0.2$--$2)\times 10^{-10}$~\cite{Bulbul_2014,Boyarsky_2014}. }
    \label{fig:nus}
\end{figure}

As a result, we obtain very stringent and model-independent constraints on the WDM masses of $> 3.6$--5.1~keV at 95\% confidence level (CL), estimated for a range of Milky-Way halo masses $M_{200} = (0.6$--$2.0)\times 10^{12} M_\odot$ (Fig.~\ref{fig:wdm}), where $M_{200}$ is defined as the enclosed mass within the radii in which the mean density is 200 times the critical density. 
We also exclude the sterile neutrino dark matter with masses lighter than 12~keV for a Milky-Way halo mass of $10^{12} M_{\odot}$ (Fig.~\ref{fig:nus}).
By assuming that only halos with masses heavier than $10^8M_\odot$ form galaxies in them, we obtain even more stringent (model-dependent) limits on the WDM masses of $>9.0$~keV for Milky-Way halo mass $10^{12} M_{\odot}$.

\section{Subhalo Models}\label{sec:model}

\subsection{Subhalo properties}

In order to estimate the number of satellites in the Milky-Way halo, we need models that describe the formation and evolution of both halos and subhalos.
The Milky-Way subhalos are characterized with the mass $m$, parameters $r_s$ and $\rho_s$ of the Navarro-Frenk-White (NFW) profile~\cite{Navarro_1997}, and the truncation radius $r_t$ beyond which the density quickly approaches to zero~\cite{Springel_2008}.
All these quantities are at the current redshift $z = 0$, after the tidal evolution of the subhalos.
In addition, some subhalos may get completely disrupted when the tidal effect strips substantial amount of masses in the outer radii such that $r_t < 0.77 r_s$~\cite{Hayashi_2003} (but see also Ref.~\cite{vandenBosch:2017ynq}).
It is therefore important to model the subhalo evolution, and relate the present quantities with those at accretion before experiencing tidal effects.

At the epoch of accretion when a halo becomes a subhalo, its density structure is completely characterized by three parameters: accretion redshift $z_a$, virial mass $m_a$, and the concentration parameter $c_a$.
In our models, we obtain all the $z = 0$ subhalo quantities ($m$, $r_s$, $\rho_s$, and $r_t$) as a function of these three parameters as we describe below.
Then the subhalo mass function, for example, can be computed as
\begin{eqnarray}
    \frac{dN_{\rm sh}}{dm} &=& \int d m_a \int dz_a \frac{d^2 N_a}{dm_a dz_a}
    \nonumber \\ && \times
    \int dc_a P(c_a|m_a,z_a) \delta_D\left(m-m(m_a,z_a,c_a)\right)
    \nonumber \\ && \times
    \Theta\left(r_t(m_a,z_a,c_a)-0.77r_s(m_a,z_a,c_a)\right),
    \label{eq:sh_mass}
\end{eqnarray}
where $\delta_D$ is the Dirac delta function and $\Theta$ is the Heaviside step function.
We can express distributions of any subhalo quantities (e.g., $r_s$, $\rho_s$, or the maximum circular volocity $V_{\rm max}$) by using the same equation and by replacing the argument of the delta function accordingly.
The number of subhalos $N_a$ accreted with mass $m_a$ at the redshift $z_a$ is encoded in $d^2N_a/(dm_a dz_a)$, which is described with the EPS formalism as discussed in Appendix~\ref{sec:A1} and~\ref{sec:A2}.
For the distribution of the concentration parameter $c_a$, we adopt the log-normal function for the mean value $\bar c_a(m_a,z_a)$ obtained in Ref.~\cite{Ludlow_2016}, and with standard deviation of $\sigma_{\log c}=0.13$~\cite{Ishiyama_2013}.
To perform the integral, we uniformly sample the subhhalo masses between $m_a=10^5 M_\odot$ and $0.1 M$ using 500 logarithmic steps, and redshift between $z_a=7$ and $0.1$ with steps of $dz=0.1$. Even though our models allow for finer resolutions, the adopted resolution reduces the computational time without affecting the results.

\subsection{Matter power spectrum} 
\label{sec:MPS}
WDM suppresses gravitational clustering and erases cosmological perturbations at scales below the WDM free-streaming length, resulting in a cutoff in the power spectrum. 
The cut-off in the matter power spectrum can be described by a transfer function, $T^2(k)$, which gives the ratio in power spectra between a WDM and CDM universe as follows~\cite{Viel_2005,Viel_2012},
\begin{equation} \label{SM:eq1}
\begin{split}
T^2(k)&\equiv P_{\rm{WDM}}(k)/P_{\rm{CDM}}(k) = \left(1+(\alpha k)^{2.24} \right) ^{-5/1.12},\\
\alpha&=0.049\left(\frac{1~ \rm{keV}}{m_{\rm{WDM}}}\right)^{1.11} \left(\frac{\Omega_{\rm{WDM}}}{0.25}\right)^{0.11} \left(\frac{h}{0.7} \right)^{1.22}, 
\end{split}
\end{equation}
where $m_{\rm WDM}$ is the WDM mass, and $P_{\rm CDM}(k)$ the linear power spectrum for CDM which we obtain from the 7-year data WMAP observations~\cite{Komatsu_2011} with corresponding cosmological parameters, $\Omega_{\rm WDM}=0.27$ and $h\equiv H_0/(100~\mathrm{km~s^{-1}})=0.7$. 

The variance of the power spectrum, $S$, is found by smoothing over a mass scale using a filter function. We adopt a ``sharp-$k$'' filter, which has been found to be well suited for truncated power spectra~\cite{Benson_2012,Lovell_2016}. The mass $M$ associated to the filter scale $R$ is, however, less well defined, and must be calibrated using simulations through a free parameter $c$, with $M=4\pi\bar{\rho}(cR)^3/3$ and $\bar{\rho}$ the average matter density of the Universe. We adopt $c=2.5$~\cite{schneider2015structure}.

While WDM particles are assumed to be produced thermally, sterile neutrinos are non-thermal at production and the power spectrum depends on their production mechanism. We adopt the Shi-Fuller~\cite{Shi_1999} mechanism in which a net lepton asymmetry value in the primordial plasma modifies the interactions between the plasma and active neutrino species, resulting in sterile neutrinos that are WDM particles as opposed to CDM production mechanisms~\cite{Petraki_2008,Kusenko_2006,Merle_2014,Lello_2015,Patwardhan_2015}. 
We consider a wide range of lepton asymmetry values for each parameter set ($m_{\nu_s}, \sin^2(2\theta)$) in order to obtain the correct dark matter abundance. 
We use the public code \texttt{sterile-dm} to obtain phase-space distributions of $\nu_s$ and $\bar{\nu}_s$~\cite{Venumadhav_2016}, and obtain the matter power spectrum using the Boltzmann code \texttt{CLASS}~\cite{lesgourgues2011cosmic}.

\subsection{Critical overdensity}
\label{sec:cr_overdens}
Dark matter halos collapse above a critical threshold $\delta_c$, which is independent of the mass scale in the CDM case, while it does depend on the mass scale for the WDM case. Considering that halo formation is suppressed at small scale, collapse becomes more difficult below a characteristic mass scale. We consider the critical overdensity as a function of both the redshift and mass $\delta_c(M,z)$~\cite{Benson_2012}. It can be described by fitting functions based on one-dimensional hydrodynamical simulations by Ref.~\cite{Barkana_2001}, which studied the collapse thresholds for WDM by modelling the collapse delay due to pressure. We adopt their fitting functions given as
\begin{equation}
\begin{split}
    \delta_{c,\rm WDM}(M,z) = \delta_{c,\rm CDM}(z) \left[h(x)\frac{0.04}{\exp(2.3 x)} \right. \\
    + \left.\left(1-h(x)\right) \exp \left(\frac{0.31687}{\exp (0.809x)} \right) \right],
\end{split}
\end{equation}
where $x=\log (M/M_J)$ and $M_J$ the effective Jeans mass of the WDM defined as,
\begin{equation}
\begin{split}
    M_J = 3.06\times 10^8 M_{\odot} \left(\frac{1+z_{\rm eq}}{3000} \right)^{1.5} \left(\frac{\Omega_m h^2}{0.15}\right)^{1/2} \left(\frac{g_X}{1.5}\right)^{-1} \\
    \times \left( \frac{m_{\rm WDM}}{1.0 \, \rm keV} \right)^{-4},
\end{split}
\end{equation}
with $z_{\rm eq}=3600(\Omega_m h_0^2/0.15)-1$ the redshift at matter-radiation equality, $g_X=1.5$ the effective number of degrees of freedom, $m_{\rm WDM}$ the thermal WDM mass, and 
\begin{equation}
    h(x)=\left(1+\exp[(x+2.4)/0.1] \right)^{-1}.
\end{equation}

Sterile neutrinos are not thermal particles, and the sterile neutrino mass needs to be converted to its corresponding thermal relic mass in order to obtain the critical overdensity. 
We convert the mass through the half-mode wavenumber $k_{hm}$, which is the wavenumber at which the transfer function is given by $T^2(k_{hm})= P_{\nu_s}/P_{\rm CDM}=0.5$.
The half-mode wavenumber for thermal WDM is given by $k_{hm}=\frac{1}{\alpha}\left(2^{1.12/5}-1 \right)^{1/2.24}$~\cite{Bose_2015}, where $\alpha$ is a function of WDM mass as given by Eq.~\ref{SM:eq1}, and can thus be compared in order to obtain the conversion.

\subsection{Subhalo evolution}
After subhalos accrete onto their host halo, they lose mass under the gravitational tidal force exerted by the host halo. Tidal stripping has not been included in many previous analytical work~\cite{schneider2015structure,Benson_2012,Menci_2012}, while it impacts on subhalo properties~\cite{Hiroshima_2018,Bartels_2015}. Following Refs.~\cite{jiang2014statistics,Hiroshima_2018}, the average mass-loss rate of dark matter subhalos is given as follows,
\begin{equation}\label{eq:massloss}
    \dot{m}(z) = -A\frac{m(z)}{\tau_{\rm dyn}(z)}\left[\frac{m(z)}{M(z)} \right]^{\zeta},
\end{equation}
where $m(z)$ is the subhalo mass, $M(z)$ is the host halo mass, and $\tau_{\rm dyn}(z)$ is the dynamical timescale.
Through Monte Carlo modeling, we find $A$ and $\zeta$ as a function of both $M(z)$ and $z$ in a WDM universe; see Appendix~\ref{sec:massloss} for more details.
This simple modeling is proven to yield results that are consistent with those of numerical N-body simulations in the CDM case~\cite{jiang2014statistics,Hiroshima_2018}.
We solve this differential equation to obtain the subhalo mass at $z = 0$, which is uniquely determined given the initial condition $m = m_a$ at $z = z_a$.

The parameters of the NFW density profile, $r_s$ and $\rho_s$, as well as the truncation radius $r_t$ also evolve as the mass loss proceeds. 
Reference~\cite{Pe_arrubia_2010} discusses the evolution of internal structure of subhalos by relating the maximum circular velocity $V_{\rm max}$ and corresponding radius $r_{\rm max}$ at accretion redshift $z_a$ and at any later redshift (see Appendix~\ref{sec:evolution} for details). We use those relations to calculate $\rho_s$ and $r_s$, and $r_t$ at $z = 0$, all as a function of $m_a$, $z_a$, and $c_a$, in order to evaluate the subhalo number with Eq.~(\ref{eq:sh_mass}).

\section{Warm dark matter constraints using satellite count}\label{sec:an}
We obtain the expected number of subhalos in the Milky-Way by integrating Eq.~(\ref{eq:sh_mass}). 
The number of subhalos depends on the Milky-Way mass, as a host with smaller mass will yield a smaller number of subhalos accreted onto it. The Milky-Way mass is uncertain and various work find values that are within the range of $M_{200}=(0.6$--$2.0)\times10^{12}M_{\odot}$ based on the latest Gaia data~\cite{Karukes_2020,Posti_2019,Eadie_2019,Fritz_2018}. 
We consider this range of the Milky-Way mass to theoretically estimate the number of satellites. 

Observationally, Ref.~\cite{Drlica_Wagner_2020} reports on ultrafaint Milky-Way satellite galaxies with the DES and PS1. They correct for the detectability of these satellites by fitting the luminosity function obtained from simulations of satellite galaxies to the DES and PS1 satellite populations. After performing this completeness correction, 270 satellite galaxies are estimated to exist within 300~kpc from the Milky-Way center and for absolute $V$-band magnitude of $M_V<0$.

The probability of obtaining the observed number of satellites $N$, for given number of satellites obtained from our models $\mu$ for each WDM parameter, is determined by the Poisson probability, $P(N|\mu)=\mu^N\exp(-\mu)/N!$. We rule out WDM models that predict too few satellite galaxies with respect to the observed number of satellites $N_{\rm obs}$ at 95\% CL as $P(>N_{\rm obs}|\mu)=\sum_{N=N_{\rm obs}}^{N=\infty}P(N|\mu)<0.05$. 
The probability can also be described by a negative binomial distribution~\cite{Boylan-Kolchin_2010}, however we find that it has no significant impact on the results of this work. 
Moreover, there might be non-Poisson effects due to for instance spatial correlations with the Magellanic clouds~\cite{Newton:2017xqg,Sales:2011id,Sales:2007hi}. These effects are expected to be minor and are not taken into account in this work.

For our canonical constraints, we assume that all subhalos host a satellite galaxy. This is, however, unlikely the case, and we could apply some satellite forming conditions.
Since imposing galaxy formation will reduce the number of satellites, which goes along the same direction as the effect of the WDM free-streaming, doing so will strengthen constraints on the WDM masses.
Therefore, with our canonical modeling, we obtain conservative constraints on WDM.

In Fig.~\ref{fig:wdm}, we show the lower limits on the WDM masses at 95\% as a function of the Milky-Way mass. In particular, the limits are stronger by a factor of 1.4 considering $M_{200}=0.6\times 10^{12}M_{\odot}$ with respect to the case of largest possible mass of $M_{200}=2\times 10^{12}M_{\odot}$. We find that we can rule out the WDM models with $m_{\rm WDM}<3.6$--5.1~keV at 95\% CL for the possible range of the Milky-Way mass. 

Next, we impose galaxy formation condition in our models. 
We adopt a model in which star formation is initiated through atomic hydrogen cooling, for which gas needs to cool down sufficiently to $\sim$10$^4$~K.
This is inefficient below halo mass (at accretion when it peaks) of $m_a \simeq 10^8 M_{\odot}$~\cite{Sawala_2015,schneider2015structure,Brooks_2014}, which we apply as a minimum mass above which we assume that all subhalos form a satellite in them.
We also show the constraints in this case in Fig.~\ref{fig:wdm} and rule out WDM mass of 9.0~keV for a Milky-Way halo mass of $10^{12}M_{\odot}$. Moreover, for Milky-Way halo mass smaller than $0.9 \times 10^{12}M_{\odot}$, all WDM mass is excluded at 95\% CL, and 
an accurate measurement of the Milky-Way halo mass could possibly become even inconsistent with both thermal WDM models and CDM.

The stellar kinematics data of the Milky-Way satellites are a powerful observable that also needs to be considered. Given that one of the observed satellites with the smallest velocity dispersion is Leo~V with $\sigma=2.3$~km~s$^{-1}$, we opt to map this line-of-sight velocity dispersion to the halo circular velocity as $V_{\rm{circ}} =\sqrt{3}\sigma=4$~km~s$^{-1}$~\cite{Simon_2019}, and use this estimate as the threshold of maximum circular velocity.
Above this threshold $V_{\rm max}$, there are 94 satellite galaxies in total; 82 estimated satellites based on the luminosity function after completeness correction, and 12 satellites that were not included in the estimate of the luminosity function. 
Higher values of $V_{\rm{max}}$ further reduce the number of luminous satellites. As shown in Fig.~\ref{fig:wdm}, we rule out $m_{\rm{WDM}}<2.2$--3.5~keV for the Milky-Way mass range of $(0.6$--$2.0)\times 10^{12}$ M$_{\odot}$.

\section{Sterile neutrino constraints}
We consider a wide range of sterile neutrino masses and mixing angles, and fix the lepton asymmetry values for each parameter set such that the correct relic density of dark matter is obtained, by adopting sterile neutrino production through to the Shi-Fuller mechanism. 
We estimate the number of satellites in the Milky Way with a best-fit Milky-Way mass of $M_{200}=10^{12}M_{\odot}$~\cite{Cautun:2019eaf}, and note that our limits depend on the Milky-Way mass as shown in Fig.~\ref{fig:wdm}. The results are shown in Fig.~\ref{fig:nus}, where the red area represents the excluded region at 95\% CL considering 270 satellites. The lower grey area is excluded, as this parameter space corresponds to the maximum allowed lepton asymmetry, which is bounded by the Big-Bang nucleosynthesis to $L_6\leq 2500$~\cite{Serpico_2005,Boyarsky_2009}. The top grey area is excluded as the mixing is too large, resulting in dark matter overproduction. 

The shape of the constraints is related to the sterile neutrino production. Non-resonant production results in warmer sterile neutrino spectral energy distributions, and thus in larger free-streaming effects. Indeed, the upper limit in Fig.~\ref{fig:nus} have stronger constraints as a result of less small-scale structure. 
Furthermore, very large lepton asymmetry delays the sterile neutrino production and yields warmer thermal distributions due to the frequent scattering between neutrinos and the plasma, and, indeed, we find stronger constraints towards the lower limit with $L_6\leq 2500$~\cite{Venumadhav_2016}.

Sterile neutrinos can decay through mixing into an active neutrino and a photon with $E_{\gamma}=m_{\nu_s}/2$, which could be observed by X-ray telescopes. Strong limits on the mixing angle are set by previous studies based on the current X-ray data~\cite{Horiuchi:2013noa, Ng:2015gfa, Perez:2016tcq, Ng:2019gch,Abazajian_2017,Caputo:2019djj,Roach_2020,foster2021deep,Cappelluti_2018,Dessert_2020,Jeltema_2015,Tamura_2015,Malyshev_2014,Aharonian_2017,Sekiya_2015,boyarsky2019surface, figueroafeliciano2015searching, Hofmann_2019,Neronov_2016}, as indicated by the hatched grey area in Fig.~\ref{fig:nus}. The black star indicates the best-fit of the unidentified $3.5$~keV line with best-fit values $m_{\nu_s}=7.1$~keV and mixing angle $\sin^2(2\theta)=7\times 10^{-11}$~\cite{Bulbul_2014,Boyarsky_2014}. 
The parameter space of sterile neutrinos is constrained to a great degree considering both X-ray and satellite constraints. In particular, sterile neutrino mass of $m_{\nu_s}\lesssim 20$~keV is excluded, which can also be confirmed with future X-ray data with all-sky X-ray instrument eROSITA~\cite{Dekker_2021,Ando_2021,Barinov_2021}.

\section{Discussion}\label{sec:disc}
The number of satellites in the Milky-Way after completeness corrections has been found to be 124 in Ref.~\cite{Newton_2018} by using a different method. Adopting 124 satellites instead of 270, however, we find that our canonical results become weaker by only $\sim$30\%.
Some of the observed satellites might be associated to the large Magellanic cloud, and they have been roughly estimated to contribute for at most 30\% to the observed satellites~\cite{Jethwa_2016}. This corresponds to 189 Milky-Way satellites, and we find $\sim$15\% weaker results.

Throughout this work, we do not incorporate baryonic effects, besides the effective prescription as a threshold mass $m_a>10^8M_\odot$, above which we assume no satellite forms in its host subhalo.
In general, including baryonic physics would reduce the number of subhalos near the center of the main halo as subhalos are more strongly disrupted in the presence of baryons~\cite{Sawala_2017}, allowing for stronger constraints. Baryonic physics could be included in our model in future work by adding a central disk in the host halo, which has been shown to reproduce the subhalo depletion well due to the additional tidal field from the central galaxy~\cite{Garrison_Kimmel_2017}. 

There are several other complementary approaches to test WDM models.
By observing the spectrum of Lyman-$\alpha$ forests in high redshift quasars~\cite{McQuinn_2016}, Ref.~\cite{Ir_i__2017} sets a lower limit of $m_{\rm WDM}>5.3$~keV at 95\% CL. Strong gravitational lensing offers another approach to detect low-mass halos in the range of $10^6$--$10^{10}M_{\odot}$. References~\cite{Hsueh_2019,Gilman_2019} find lower limits of $m_{\rm WDM}>5.58$~keV and $m_{\rm WDM}>5.2$~keV at 95\% CL, respectively. Other independent approaches yield similar constraints (e.g., Refs~\cite{Banik:2019smi, Shirasaki:2021orc}).

\subsection{Comparison with results in the literature}
We compare our limits with previous results~\cite{schneider2015structure,Cherry_2017,Newton_2021,Kennedy_2014,Nadler_2021,Lovell_2014,Polisensky_2011} in Fig.~\ref{fig:wdm}. \\
Among them, Ref.~\cite{Nadler_2021} (triangle) adopted an empirical model for the galaxy-halo connection and obtained stringent limits of $m_{\rm WDM}>6.5$~keV. This is the result of inferring the peak halo mass above which halos host galaxies as one of the free parameters.
It is similar to our approach using the $m_a>10^8M_\odot$ galaxy-formation threshold, for which we obtain comparable constraints.
We believe that our models are more flexible because they are not limited by numerical resolutions, and also allow setting model-independent constraints. 

The results of Ref.~\cite{schneider2015structure} is shown as the plus symbol, where a semi-analytical method is presented based on the EPS formalism and calibrated to numerical simulations, adopting the satellite forming condition on the minimum subhalo mass of $m_a=10^8M_{\odot}/h$. We improve on their work as we include tidal stripping effects for the first time, which results in fewer surviving satellites, and consider a larger data set. This also applies to results on the sterile neutrino constraints from Ref.~\cite{Cherry_2017} which adopted the same semi-analytical model from Ref.~\cite{schneider2015structure}, and find a constraint on the sterile neutrino mass of $m_{\nu_s}\gtrsim 6.5$~keV, which corresponds to the WDM mass of $m_{\rm WDM}\simeq 2.5$~keV, indicated as the circle in Fig.~\ref{fig:wdm}. 

Moreover, the dashed line represents the results from Ref.~\cite{Newton_2021}, which similarly estimates the number of satellites based on the EPS method and calibrated to N-body simulations.
As a second analysis, Ref.~\cite{Newton_2021} incorporates galaxy formation processes by using the semi-analytical model for galaxy formation \texttt{GALFORM}. Marginalizing over the uncertainties in the Milky-Way halo mass they rule out $m_{\rm WDM}<3.99$~keV, while they rule out $m_{\rm WDM}<2.02$~keV without incorporating a galaxy formation model. 

The dotted line represents the results of Ref.~\cite{Kennedy_2014}, where they obtain the limits based on the EPS formalism combined with \texttt{GALFORM}. Galaxy formation models are more physically motivated, but the results depend on the choice of various parameters. Indeed, depending on the choice of the main parameter of reionization, Ref.~\cite{Kennedy_2014} finds constraints that vary between $m_{\rm WDM}<2.5$~keV and being all ruled out for the Milky-Way halo mass of $2\times10^{12}M_{\odot}$. 

We also show the limits based on N-body simulations, such as Ref.~\cite{Lovell_2014} (star) and Ref.~\cite{Polisensky_2011} (cross). These numerical simulations are, however, limited by the numerical resolution, and moreover, Ref.~\cite{Lovell_2014} probes a maximum dark matter mass of up to $m_{\rm WDM}\leq 2.3$~keV, while Ref.~\cite{Polisensky_2011} probes five dark matter masses between $m_{\rm WDM}=1$~keV and 5~keV. We test our model with respect to the results from N-body simulations of Ref.~\cite{Lovell_2014}, as discussed in the following section.

In the following we discuss consequences of adopting different peak mass thresholds and fitting function of the subhalo mass function~\cite{Lovell_2014} often used in the literature. 

\subsection{Threshold on subhalo peak mass}
\begin{figure}[ht!]
    \centering
    \includegraphics[width=0.48\textwidth]{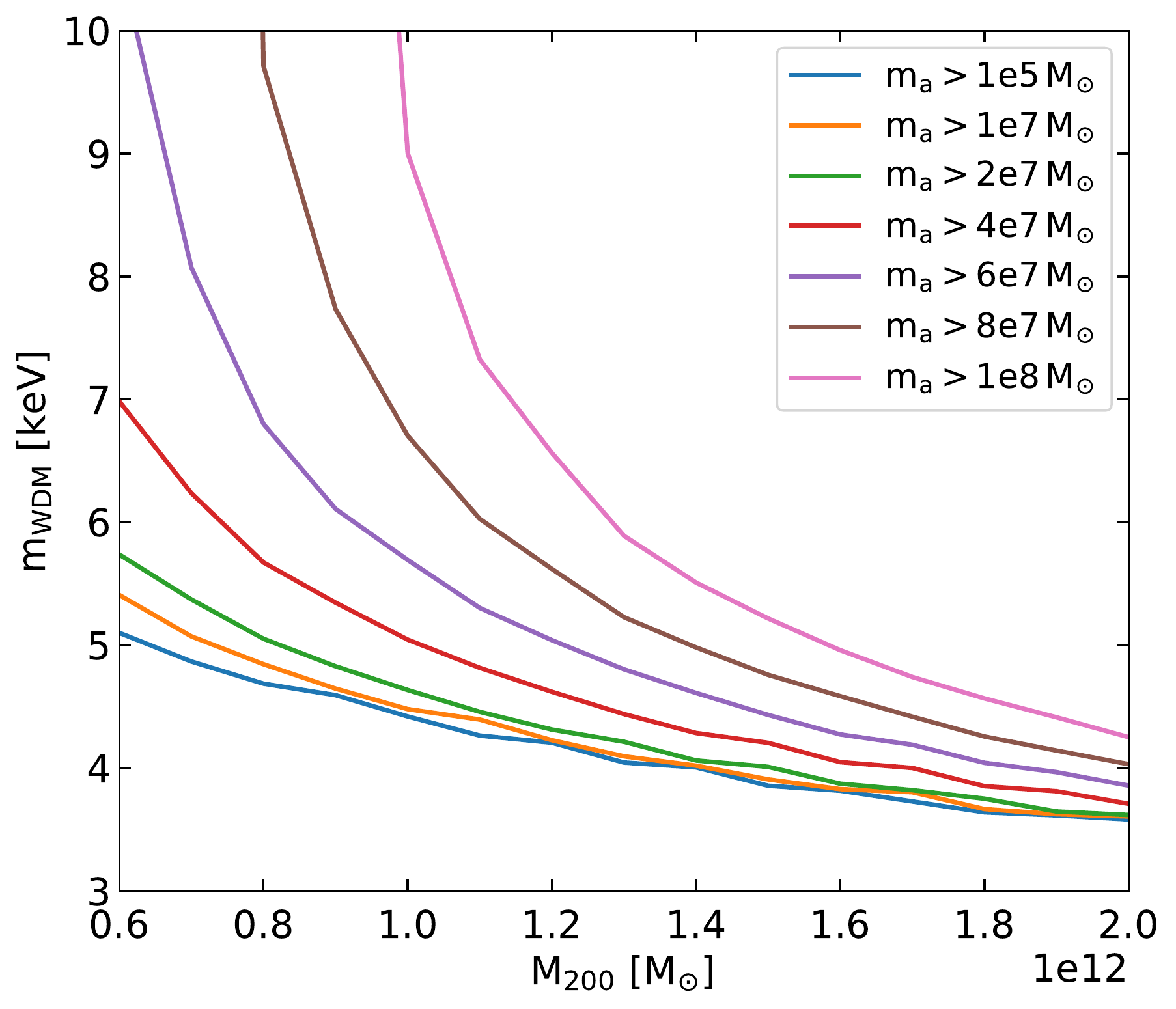}
\caption{Lower limits at 95\% CL on the WDM mass as a function of the Milky-Way mass considering different thresholds for galaxy formation on the peak subhalo mass $m_a$.}
    \label{fig:Mpeak}
\end{figure}

In our canonical model, we adopt a minimum subhalo mass at accretion (peak mass) of $m_a=10^5 M_{\odot}$. In order to effectively set a satellite forming condition, we impose a threshold on the peak mass of $m_a>10^8 M_{\odot}$, as has been done in the previous work~\cite{schneider2015structure,Cherry_2017}. Note, however, that the threshold on the peak mass is model-dependent, and we therefore show in Fig.~\ref{fig:Mpeak} the effect of adopting different thresholds on the peak mass. We find that for a Milky-Way halo mass of $M_{200}=10^{12}M_{\odot}$, the constraints can vary between 4.5~keV and completely excluded by adopting a threshold on the peak mass between $10^7 M_\odot$ and $10^8M_{\odot}$. 

\subsection{Comparison with N-body simulations}\label{sec:Numerical}
\begin{figure}[hbt!]
\centering
{\includegraphics[width=0.98\linewidth]{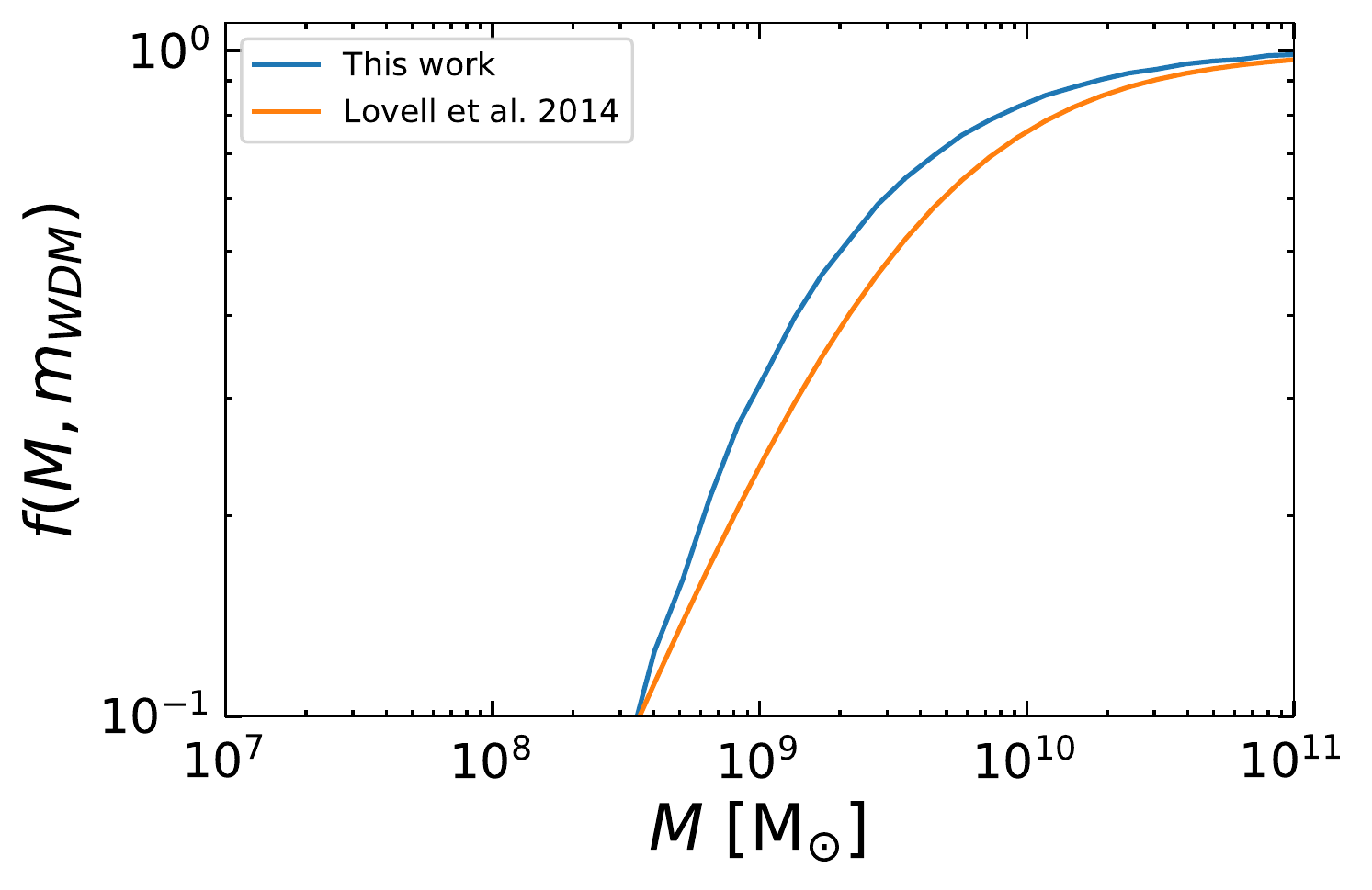}}\hfill
{\includegraphics[width=0.98\linewidth]{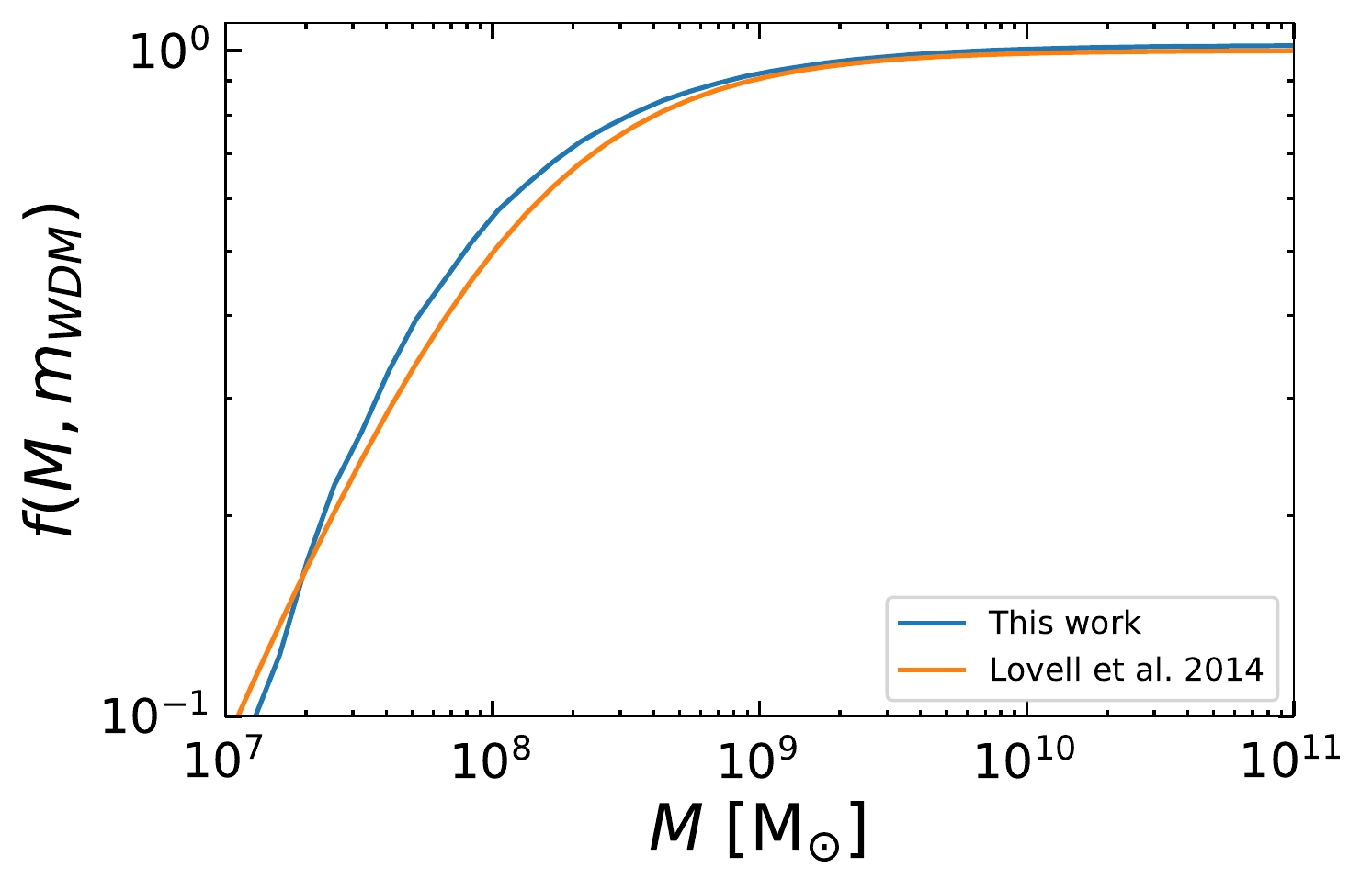}}
\caption{Subhalo mass function suppression of WDM with respect to CDM adopting $m_{\rm WDM}=2.3$~keV (upper) and $m_{\rm WDM}=6.5$~keV (bottom). The subhalos mass $M$ is the virial mass before the tidal mass loss. }
\label{fig:SHMF}
\end{figure}
In order to test our models with numerical simulations, we first compare the suppression of the subhalo mass functions due to WDM with respect to CDM, defined as
\begin{equation}\label{eq:SHMF}
    f(M,m_{\rm WDM})=\frac{(dN_{\rm sh}/dM)_{\rm WDM}}{(dN_{\rm sh}/dM)_{\rm CDM}},
\end{equation}
where $M$ is the subhalo mass {\it before} mass loss. Based on high-resolution cosmological N-body simulations, Ref.~\cite{Lovell_2014} find a functional form for Eq.~(\ref{eq:SHMF}) by fitting the simulations with $m_{\rm WDM}= 2.3$, 2.0, 1.6 and 1.5~keV. We compare the subhalo mass function suppression in Fig.~\ref{fig:SHMF} for $m_{\rm WDM}=2.3$~keV with the Milky-Way mass of $M_{200}=1.9\times 10^{12}M_{\odot}$ (left), and 6.5~keV with $M_{200}=1.4\times 10^{12}M_{\odot}$ (right). We find a weaker suppression, but it differs at most by only a few tens of percent. The fitting function for subhalo mass before tidal stripping should, however, not be confused with the actual subhalo mass after it evolved due to tidal stripping, whereas there appears to be some level of confusion in the literature.

\begin{figure}[ht!]
\centering
{\includegraphics[width=0.98\linewidth]{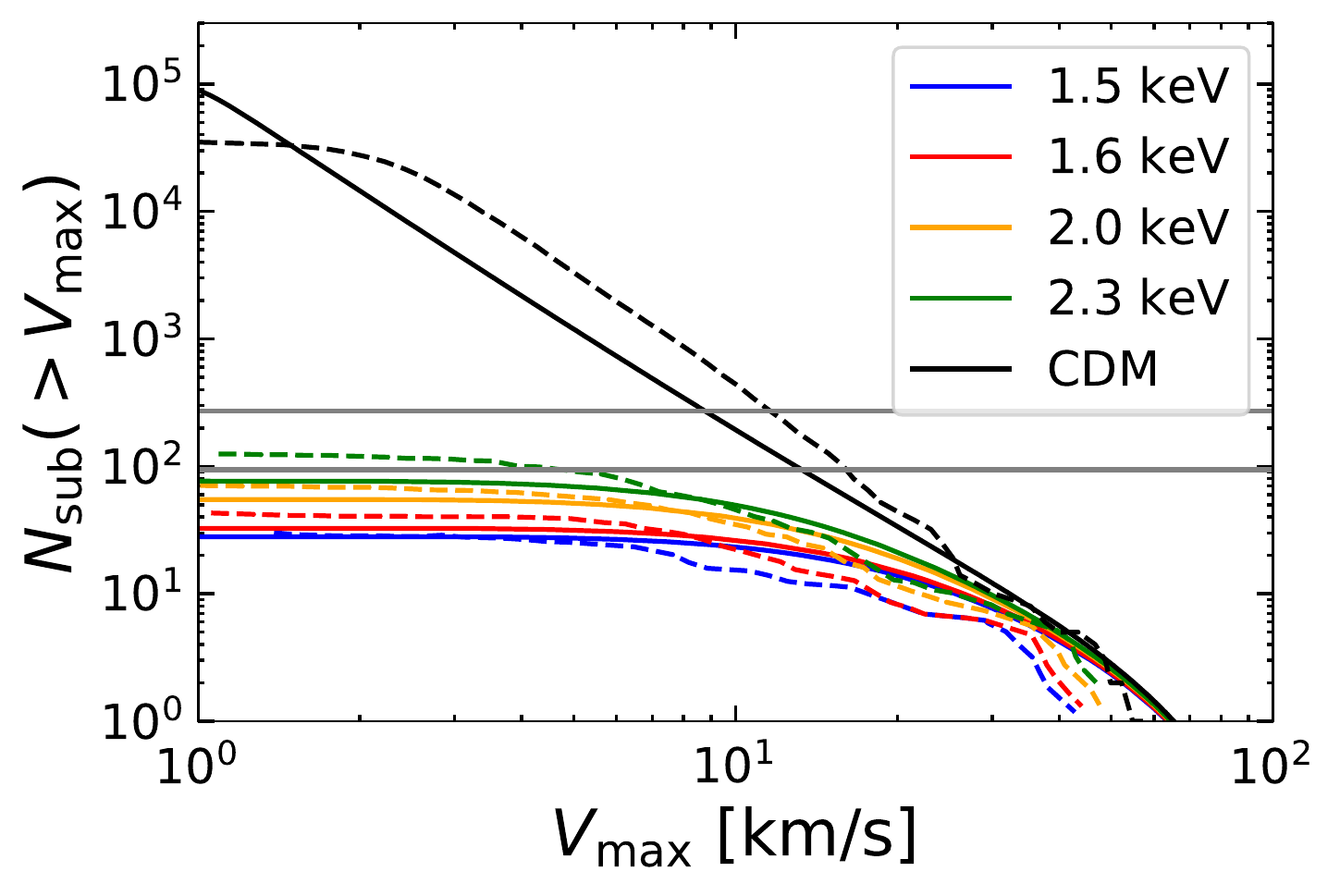}}\hfill
{\includegraphics[width=0.98\linewidth]{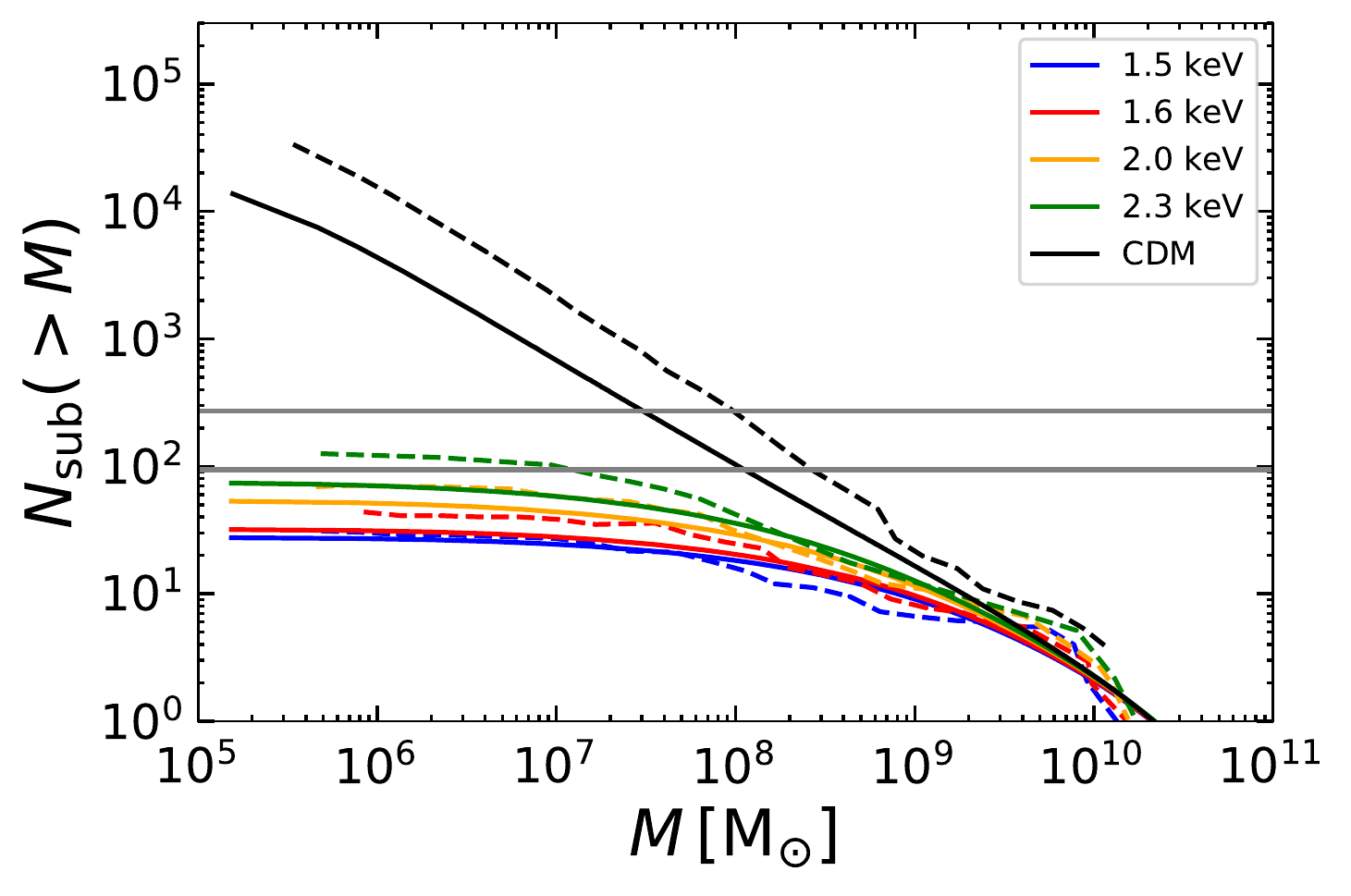}}
\caption{Cumulative maximum circular velocity (upper) and subhalo mass (bottom) function for the number of subhalos in the Milky Way with mass $M_{200}=1.8\times 10^{12}M_{\odot}$. Dashed lines are the results of N-body simulations obtained from Ref.~\cite{Lovell_2014} and the solid lines are our results. }
\label{fig:N_lovell}
\end{figure}

Next, we compare the cumulative maximum circular velocity and subhalo mass functions for subhalos {\it after} tidal stripping with the results from Ref.~\cite{Lovell_2014}, as shown in Fig.~\ref{fig:N_lovell}. The dashed lines are the results from Ref.~\cite{Lovell_2014} and solid lines are our results. We find an underestimation with respect to the N-body simulation results; they differ by a factor of 1.6 at most for WDM, and a factor of 2.5 for CDM. Moreover, we also show two horizontal grey solid lines, that correspond to the number of observed satellites that we use in our analysis, $N_{\rm sat}=270$ and 94. 

The simulations by Ref.~\cite{Lovell_2014} are based on the Aquarius simulation for CDM~\cite{Springel_2008}.
It was noted that the subhalo mass functions from Aquarius were found to be larger than the results of many other similar N-body simulations by a factor of a few. 
The cause of this discrepancy is not completely understood, but might be related to the halo-to-halo variance, as one cannot simulate very many Milky-Way-like halos with simulations like Aquarius that are tuned to have much greater numerical resolutions.
In any case, if the WDM simulations were implemented based on other CDM runs, we expect that the degree of discrepancy that we see in Fig.~\ref{fig:N_lovell} is much smaller.
Thus, together with Fig.~\ref{fig:SHMF}, we believe that our models based on \texttt{SASHIMI} predict subhalo quantities for both before and after the tidal mass loss in a WDM cosmology.

As far as we are aware, there is no convenient fitting function like the one proposed in Ref.~\cite{Lovell_2014} for the subhalo mass functions after the tidal mass loss.
We therefore encourage the community to use \texttt{SASHIMI} instead, whenever the subhalo properties after the tidal evolution are the quantities in question.


\section{Conclusion}
The satellite number counts provide one of the most reliable and stringent constraints on long-debated WDM models, for which our semi-analytical approach that combines halo formation with tidal evolution enables predictions for a wide range of both WDM and Mikly-Way halo masses.
We make the numerical codes, Semi-Analytical SubHalo Inference ModelIng (\texttt{SASHIMI}), publicly available for both CDM\footnote{\url{https://github.com/shinichiroando/sashimi-c}} and WDM.\footnote{\url{https://github.com/shinichiroando/sashimi-w}}  
\texttt{SASHIMI} provides a flexible and versatile platform for computing subhalo quantities and the constraints obtained with it are one of the best and most robust, being independent of physics of galaxy formation and free from numerical resolution and the Poisson noise.
By comparing the latest satellite number counts obtained by the DES and PS1 surveys, we exclude the WDM masses for a wide range of Milky-Way halo mass, and find lower bounds of $m_{\rm WDM}>4.4$~keV at 95\% CL for Milky-Way halo mass of $10^{12}M_{\odot}$, independent of galaxy formation physics. By adopting a galaxy-formation condition, we find that the limits significantly improve to $m_{\rm WDM}>9.0$~keV. Moreover, we obtain limits on sterile neutrino masses of $m_{\nu_s}>12$~keV, and combined with current X-ray limits, $m_{\nu_s}>20$~keV.
Our results thus show that there remain smaller rooms for warm particles such as thermal WDM and keV sterile neutrinos to be a dominant dark matter candidate.

\acknowledgments{We are grateful to Nagisa Hiroshima for helpful discussions, and Andrew Benson and Mark Lovell for comments on the manuscript. We also thank Aaron Ludlow for sharing the codes on the mass-concentration-redshift relation. The work of SA was supported by JSPS/MEXT KAKENHI Grant Numbers JP17H04836, JP20H05850, and JP20H05861. CAC acknowledges the support by the Dutch Research Council (NWO Veni 192.020).}

\appendix
\section{Mass Accretion History of the host halo~\label{sec:A1}}
\begin{figure}[ht!]
    \centering
    \hskip7.mm
    \includegraphics[width=0.48\textwidth]{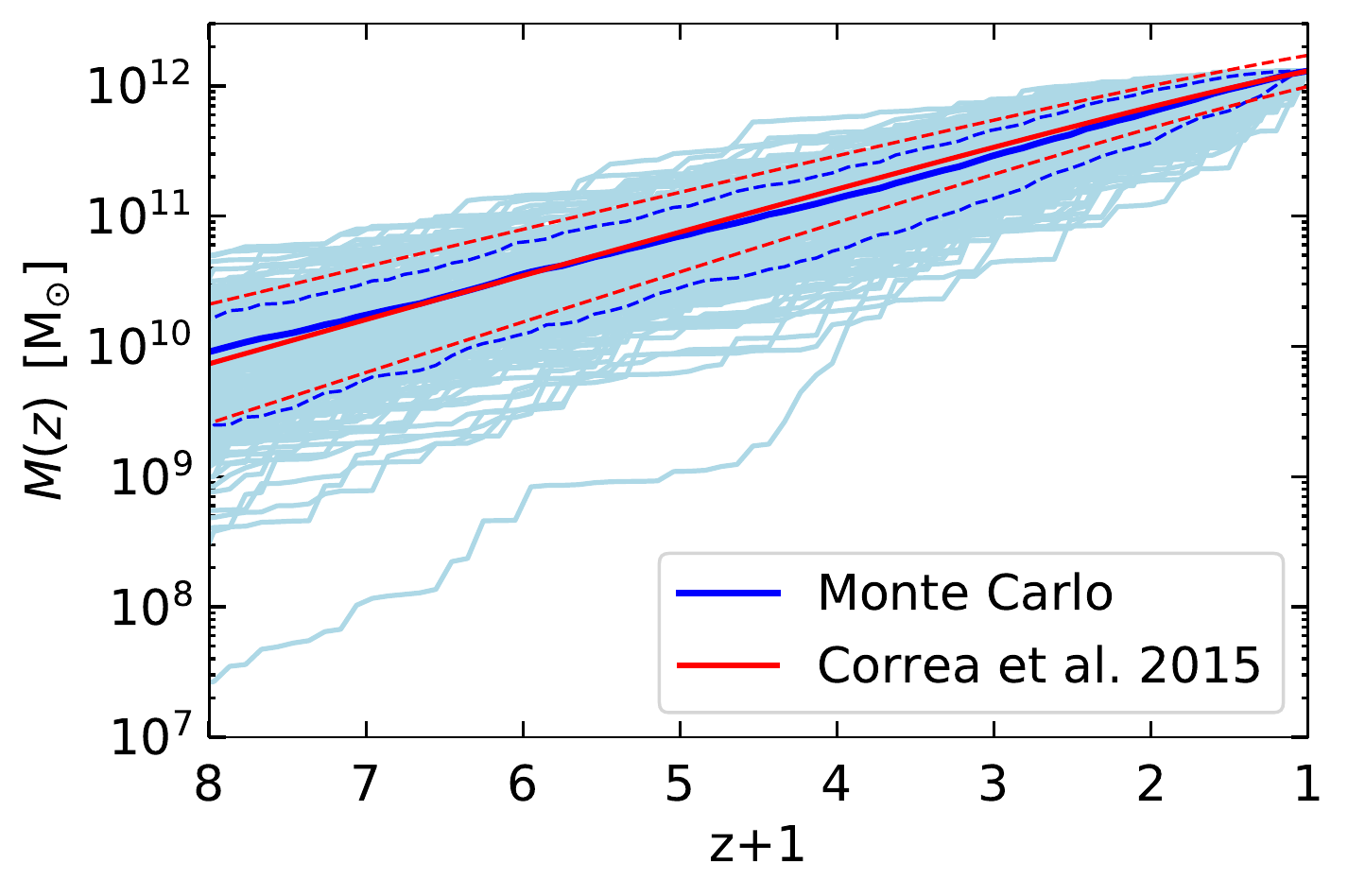}
\caption{Comparison of mass accretion history from 200 Monte Carlo runs with corresponding mean (blue solid line) and standard deviation (blue dashed line), with the mean mass accretion history obtained from Ref.~\cite{Correa_2015_I} for a CDM universe (red solid line) and its standard deviation (red dashed line). A WDM mass of $m_{\rm WDM}=1.5$~keV is assumed with host halo $M_0=1.3\times 10^{12} M_{\odot}$ at $z=0$.}
    \label{fig:MAH}
\end{figure}

The mass function of WDM haloes is suppressed with respect to CDM haloes below the cut-off scale $M_s \sim 10^8 M_{\odot}$~\cite{Barkana_2001}. Meanwhile, the mean mass accretion history (MAH) above the cut-off scale shows almost no difference between WDM and CDM haloes~\cite{Benson_2012}. This allows to adopt analytical expressions for the MAH that are derived for a CDM universe. In order to confirm this, we use the fraction of halo mass ($M_1,z_1$) that is in progenitor halo mass at some later redshift ($M_2,z_2$), where we use Model II for the fraction $f(S_2,\delta_2|S_1,\delta_1)$ as defined in Ref.~\cite{Yang_2011}. By Monte Carlo method, we simulate 100 host halo masses with $M_0=1.3\times 10^{12} M_{\odot}$ for WDM mass $m_{\rm WDM}=1.5$~keV between redshift $z=0$ and 10. We show the obtained MAH in Fig.~\ref{fig:MAH}, where the light blue lines show the individual MAH for each run, and the blue solid and dashed lines correspond to the mean and the standard deviation respectively. We compare the results with the analytical expressions for the MAH from Ref.~\cite{Correa_2015_I} obtained for a CDM universe, indicated by the red solid and dashed lines, representing the MAH and standard deviation $\sigma_{\log M_a}=0.12-0.15 \log (M_a / M_0)$ respectively. They show similar behavior and we therefore adopt the analytical expressions from Ref.~\cite{Correa_2015_I} for the MAH as follows,
\begin{equation}\label{eq:MAH}
    M(z)=M_0(1+z)^{\alpha}\exp(\beta z),
\end{equation}
with parameters $\alpha$ and $\beta$

\begin{equation}
\begin{split}
    \beta &= -f(M_0),\\
    \alpha &= \left[1.686(2/\pi)^{1/2} \left.\frac{dD}{dz}\right|_{z=0} +1  \right] f(M_0),
\end{split}    
\end{equation}
and with
\begin{equation}
\begin{split}
    f&(M_0) = \left[S(M_0/q) - S(M_0) \right]^{-1/2},\\
    q &= 4.137 \Tilde{z}_f^{-0.9476},\\
    \Tilde{z}&_f = -0.0064(\log_{10} M_0)^2 + 0.0237(\log_{10} M_0) + 1.8827,
\end{split}    
\end{equation}
where $D(z)$ is the linear growth factor and $S(M)\equiv \sigma^2(M)$ the variance of the matter density smeared over scales corresponding to $M$ with some filter function. The MAH can be further generalized to obtain the mass $M(z)$ at redshift $z$, which had a mass of $M(z_i)$ at redshift $z_i$~\cite{Correa_2015_III}:
\begin{equation}
    M(z) = M(z_i)(1+z-z_i)^{\alpha}\exp(\beta (z-z_i)). 
\end{equation}

\section{Mass Accretion History of the subhalo~\label{sec:A2}}
With the expression for the MAH of the main branch, we can obtain the subhalo mass function accreted at a certain redshift onto the main branch, and we follow the model by Ref.~\cite{Yang_2011}. 
The distribution of the masses $m_a$ and redshifts $z_a$ of subhalos that accreted onto the main branch of host halo that would evolve to $M_0$ at $z=0$, is given by~\cite{Yang_2011}
\begin{equation}
    \frac{d^2 N_a}{d \ln m_a dz_a}= \mathcal{F}\left(s_a,\delta_a | S_0,\delta_0;\overline{M}_a\right)
    \left|\frac{ds_a}{dm_a}\right|\left|\frac{d\overline{M}_a}{dz_a}\right|,
\end{equation}
where $N_a$ are the number of subhalos of mass $m_a$ at accretion redshift $z_a$, $\overline{M}_a\equiv \overline{M}(z_a)$ is the mean mass of the main branch at accretion given by Eq.~(\ref{eq:MAH}), $\mathcal{F}\left(s_a,\delta_a|S_0,\delta_0;\overline{M}_a\right)$ is the mass fraction in subhalos of mass $m_a$ at $z_a$ that accreted when the host was in the mass range $[\overline{M}_a-d\overline{M}_a,\overline{M}_a]$ as described hereafter. Furthermore, $s_a\equiv \sigma^2(m_a)$ and $S_0\equiv \sigma^2(M_0)$ are the variances of the density fluctuation, and $\delta_a \equiv \delta_c(z_a)$ and $\delta_0\equiv \delta_c(0)$, where $\delta_c$ is the threshold value of the gravitational collapse above which the overdense region is assumed to have collapsed to form a virialized halo. The variance and critical overdensity differ in the case of a WDM universe with respect to the standard CDM case in which the EPS formalism was described, and we incorporate adjustments as discussed in the sections~\ref{sec:MPS} and \ref{sec:cr_overdens}. 

We assume that the probability distribution of the host, $M_a$, at accretion redshift $z_a$ follows a log-normal distribution with logarithmic mean value of $\overline{M}_a=M(z_a)$ [Eq.~(\ref{eq:MAH})], and logarithmic dispersion $\sigma_{\log M_{a}}=0.12-0.15\log ({M_a}/{M_0})$. 
Following Ref.~\cite{Yang_2011}, $\mathcal{F}$ is defined as
\begin{equation}
\begin{split}
    \mathcal{F}\left(s_a,\delta_a|S_0,\delta_0;\overline{M}_a\right) 
    = \int \Phi\left(s_a,\delta_a|S_0,\delta_0;\overline{M}_a\right)\\
    \times P(M_a|S_0,\delta_0)dM_a,
\end{split}
\end{equation}
where 
\begin{equation}
\begin{split}
    \Phi\left(s_a,\delta_a|S_0,\delta_0;\overline{M}_a\right) = 
    \left[\int_{S(m_{\rm max})}^{\infty} F\left(s_a,\delta_a|S_0,\delta_0;M_a\right) \right]^{-1}\\
    \times \left\{ 
  \begin{array}{ c l }
    F\left(s_a,\delta_a|S_0,\delta_0;M_a\right), & \quad \textrm{if } m_a\leq m_{\rm max}, \\
    0 ,                & \quad \textrm{otherwise},
  \end{array}
\right.
\end{split}
\end{equation}

\begin{equation}
\begin{split}
    F(s_a,\delta_a | S_0,\delta_0;M_a) d \ln \delta_a 
    = \frac{1}{\sqrt{2 \pi}} \frac{\delta_a-\delta_M}{(s_a-S_M)^{3/2}}\\
    \times \exp\left[-\frac{(\delta_a-\delta_M)^2}{2(s_a-S_M)} \right],
\end{split}    
\end{equation}
where $m\leq m_{\rm max}\equiv \min[M_a,M_0/2]$ as the subhalos accreting onto the main branch of its merger tree, and ($S_M$, $\delta_M$)
are defined at redshift value for which $M=M_{\rm max}=\min[M_a+m_{\rm max},M_0]$, as the main branch will increase its mass due to accretion of $m_a$.

\section{Mass-loss rate}
\label{sec:massloss}
We adopt a toy model to describe the subhalo mass loss following Refs.~\cite{jiang2014statistics,Hiroshima_2018}, in which all mass is assumed to be lost during the first orbital period within the host halo, in order to find the parameters for $A$ and $ \zeta$ of Eq.~5.
By Monte Carlo method, we obtain the mass-loss rate by considering host masses in the range $M=[10^{-6}, 10^{16}]M_{\odot}$ and redshift range $z=[0,7]$, and fit the values of $A$ and $ \zeta$. We find the following fitting functions,
\begin{equation}
\begin{split}
\log A = \left[ -0.0019 \log \left( \frac{M(z)}{M_{\odot}} \right) + 0.045 \right]z \\
+ 0.0097 \log \left(\frac{M(z)}{M_{\odot}} \right) -0.31,
\end{split}
\end{equation}

\begin{equation}
\begin{split}
    \zeta = \left[ -0.000056 \log \left( \frac{M(z)}{M_{\odot}} \right) + 0.0014 \right]z \\
    + 0.00033 \log \left(\frac{M(z)}{M_{\odot}} \right) - 0.0081 . 
\end{split}
\end{equation}
Reference~\cite{Hiroshima_2018} find slightly different fitting functions in the case of CDM. In both cases, however, $A$ and $\zeta$ only weakly depend on the host mass and redshift.

\section{Subhalo structure before and after tidal stripping~\label{sec:evolution}}
We assume that the subhalos follow a NFW profile with a sharp drop at the truncation radius $r_t$ as follows,
\begin{equation}
    \rho(r) = 
    \begin{cases}
      \rho_s r_s^3 / \left[r(r+r_s) \right]^2, & \mbox{for }r \leq r_t,\\
      0, & \mbox{for }r>r_t,
    \end{cases}    
\end{equation}
where $r_s$ is the scale radius and $\rho_s$ the characteristic density which is obtained as
\begin{equation}
    \rho_s = \frac{m}{4\pi r_s^3 f(c_{\rm vir})},
\end{equation}
with $f(c)=\ln(1+c)-c/(1+c)$ and $c_{\rm vir}=r_{\rm vir}/r_s$ the virial concentration parameter. The parameters $r_s$ and $\rho_s$ are related to the maximum circular velocity, $V_{\rm max}$, and the corresponding radius $r_{\rm max}$ as follows,
\begin{equation}
\begin{split}
    r_s &= \frac{r_{\rm max}}{2.163},\\
    \rho_s &= \frac{4.625}{4\pi G} \left(\frac{V_{\rm max}}{r_s}\right)^2.
\end{split}
\end{equation}

The evolution of subhalos before and after tidal stripping is found in Refs.~\cite{Pe_arrubia_2010,Hiroshima_2018,Ando_2019} by relating the maximum circular velocity $V_{\rm max}$ and corresponding radius $r_{\rm max}$ at accretion redshift $z_a$ and at any later redshift $z_0$,
\begin{equation}
\begin{split}
    \frac{V_{\rm max,0}}{V_{\rm max,a}}=\frac{2^{0.4}(m_0/m_a)^{0.3}}{(1+m_0/m_a)^{0.4}},\\
    \frac{r_{\rm max,0}}{r_{\rm max,a}}=\frac{2^{-0.3}(m_0/m_a)^{0.4}}{(1+m_0/m_a)^{-0.3}},
\end{split}
\end{equation}
where $m_0/m_a$ is the mass ratio between the subhalo after and before tidal stripping. The truncation radius can then be found by relating $r_{\rm max}$ and $V_{\rm max}$ to the scale radius $r_s$ and characteristic density $\rho_s$ for the NFW profile. Using these relations, the subhalo mass after tidal stripping can be obtained by solving
\begin{equation}
    m_0=4\pi\rho_{s,0} r_{s,0}^3 f\left(\frac{r_{t,0}}{r_{s,0}} \right).
\end{equation}
We assume that subhalos with ratios $r_{t,0}/r_{s,0}<0.77$ do not survive due to tidal disruption and remove them from further calculations~\cite{Hayashi_2003}. 

\section{Mass-concentration-redshift relation \label{sec:conc}}
The concentration of dark matter halos depends on their MAH, and, as WDM halos form later than CDM halos, they are expected to have a lower concentration due to a lower background density at later time. In particular, the concentration in the WDM case peaks at a mass scale related to its truncation scale, in contrary to the CDM case that has monotonic relations between mass, concentration and redshift. 
We adopt the mean concentration-mass-redshift relation, $\bar c_a(m_a,z_a)$ obtained in Ref.~\cite{Ludlow_2016}, which is inferred from the MAH and can be applied to any WDM model.
The concentration parameter $c_a$ is then drawn by following the log-normal distribution $P(c_a|m_a,z_a)$ around this mean $\bar c_a$.

\bibliography{references}

\end{document}